\documentclass[twocolumn,english,superscriptaddress]{revtex4-1}
\usepackage[T1]{fontenc}
\usepackage{amsfonts}
\usepackage{amsmath}
\usepackage[latin9]{inputenc}
\usepackage{textcomp}
\usepackage{amstext}
\usepackage{subscript}
\usepackage{graphics}
\usepackage{graphicx}
\usepackage{float}

\makeatletter

\newcommand{\lyxmathsym}[1]{\ifmmode\begingroup\def\b@ld{bold}
  \text{\ifx\math@version\b@ld\bfseries\fi#1}\endgroup\else#1\fi}

\usepackage{hyperref}
\usepackage{upgreek,textgreek}
\hypersetup{colorlinks=true,linkcolor=blue,citecolor=blue,urlcolor=blue}
\makeatother

\usepackage{babel}
\begin{document}
\title{Multiband effects in thermoelectric and electrical transport properties of  kagome superconductors \textit{A}V\textsubscript{3}Sb\textsubscript{5} (\textit{A} = K, Rb, Cs)}
\author{Xinrun Mi}
\thanks {These authors contributed equally to this work.}
\affiliation{Low Temperature Physics Lab, College of Physics \& Center of Quantum
Materials and Devices, Chongqing University, Chongqing 401331, China}
\author{Wei Xia \textcolor{blue}{\textsuperscript{*}}}

\affiliation{School of Physical Science and Technology, ShanghaiTech University, Shanghai 201210, China}
\affiliation{
 ShanghaiTech Laboratory for Topological Physics, Shanghai 201210, China}
\author{Long Zhang\textcolor{blue}{\textsuperscript{*}}}
\affiliation{Low Temperature Physics Lab, College of Physics \& Center of Quantum
Materials and Devices, Chongqing University, Chongqing 401331, China}
\author{Yuhan Gan}
\affiliation{Low Temperature Physics Lab, College of Physics \& Center of Quantum
Materials and Devices, Chongqing University, Chongqing 401331, China}
\author{Kunya Yang}
\affiliation{Low Temperature Physics Lab, College of Physics \& Center of Quantum
Materials and Devices, Chongqing University, Chongqing 401331, China}
\author{Aifeng Wang}
\affiliation{Low Temperature Physics Lab, College of Physics \& Center of Quantum
Materials and Devices, Chongqing University, Chongqing 401331, China}
\author{Yisheng Chai}
\affiliation{Low Temperature Physics Lab, College of Physics \& Center of Quantum
Materials and Devices, Chongqing University, Chongqing 401331, China}
\author{Yanfeng Guo}
\email{guoyf@shanghaitech.edu.cn}
\affiliation{School of Physical Science and Technology, ShanghaiTech University, Shanghai 201210, China}
\author{Xiaoyuan Zhou}
\email{xiaoyuan2013@cqu.edu.cn}

\affiliation{Low Temperature Physics Lab, College of Physics \& Center of Quantum
Materials and Devices, Chongqing University, Chongqing 401331, China}
\author{Mingquan He}
\email{mingquan.he@cqu.edu.cn}

\affiliation{Low Temperature Physics Lab, College of Physics \& Center of Quantum
Materials and Devices, Chongqing University, Chongqing 401331, China}
\date{\today}

\begin{abstract}
We studied the effects of multiband electronic structure on the thermoelectric and electrical transport properties in the normal state of  kagome superconductors \textit{A}V\textsubscript{3}Sb\textsubscript{5} (\textit{A} = K, Rb, Cs).  In all three members, the multiband nature is manifested by sign changes in the temperature dependence of the Seebeck and  Hall resistivity, together with sublinear response of the isothermal Nernst and Hall effects to external magnetic fields in the charge ordered state. Moreover, ambipolar transport effects appear ubiquitously in all three systems, giving rise to sizable Nernst signal.  Finally, possible origins of the sign reversal in the temperature dependence of the Hall effect are discussed.
 
\end{abstract}

\maketitle
\section{Introduction}
Despite its simple structure,  the kagome lattice is a novel playground to study the interplay between geometrical frustration, band topology and electronic correlations \cite{Syozi1951,Balents2010,Ortiz2020,Jiang_kagome,Neupert2022}.  Of particular interest is the newly discovered V-based kagome superconductors  $A$V$_3$Sb$_5$ ($A=$ K, Rb, Cs) \cite{Ortiz2019,Ortiz2020,Ortiz2021,Yin2021}. The kagome net in  $A$V$_3$Sb$_5$ is formed by V atoms with $d$-orbital characteristics, leading to multiple van Hove singularities and non-trivial band topology near the Fermi level \cite{Ortiz2019,Ortiz2020,Li2021,Nakayama2021,Wang2021CDW,Liu2021,Ortiz2021Cs,Hu_2022}. A charge density wave (CDW) state appears in all three members, followed by  a superconducting instability at lower temperatures \cite{Ortiz2019,Ortiz2020,Ortiz2021,Yin2021,Jiang2021a,Liang2021CDW,Zhao2021a,Chen2021rotonpair,Xu2021,Li2021a,Shumiya2021,Wang2021CDW}.  Both the charge order and the superconducting ground state are likely unconventional, and compete with each other in a complex manner under pressure \cite{Li2021NodalSA,Chen2021,Du2021,Chen2021a,Yu2021a,Zhu2021}. Despite the absence of long-range magnetic orders, the charge order possibly breaks time-reversal symmetry, as evidenced by the appearance of electronic chirality \cite{Jiang2021a,Wang2021CDW,Shumiya2021}, a giant anomalous Hall effect (AHE) \cite{Yang2020,Yu2021,Zheng2021gate}, an anomalous Nernst effect \cite{Chendong2021}, an anomalous thermal Hall effect \cite{Zhou2021}, and an internal magnetic field in the CDW state \cite{Mielke2021,Yu2021b}. The unconventional aspects of the superconducting order are supported by the observations of residual thermal conductivity \cite{Li2021NodalSA}, Majorana bound states \cite{Liang2021CDW}, pair density wave \cite{Chen2021rotonpair}, charge-4$e$ and charge-6$e$ flux quantization \cite{Ge2022}. Additionally, a rotational symmetry breaking is  observed in the CDW and superconducting phases, suggesting intertwined charge, electronic nematic and superconducting orders \cite{Xiang2021,Nie2022}. Investigation of the interplay between these intertwined orders is the current focus of the $A$V$_3$Sb$_5$ kagome family. The three members, KV$_3$Sb$_5$, RbV$_3$Sb$_5$ and CsV$_3$Sb$_5$, share various similarities. Track down the common features across all members is essential to study the underlying generic physics in $A$V$_3$Sb$_5$. 

One of the common properties shared by all series in $A$V$_3$Sb$_5$ is a multiband fermiology \cite{,Ortiz2020,Ortiz2019,Li2021,Nakayama2021,Wang2021CDW,Liu2021,Ortiz2021Cs,Hu_2022,Tan2021,Zhao2021a,Luo2022}. The electronic structure possesses a quasi-2D fermiology, which mainly consists of a circular electron sheet and a hexagonal hole sheet centered at the $\varGamma$ point, a triangular electron band and a triangular hole band around the $K$ point, a tiny electron pocket around the $L$ point, and a small hole pocket at the $M$ point \cite{Tan2021,Li2021,Luo2022}.    The charge ordering mainly affects the bands near the zone boundary, and the CDW gap appears to be largest at the $M$ point \cite{Tan2021,Luo2022,Ortiz2021Cs}.   It is very likely that the unconventional properties are also governed by the bands and van Hove singularities near the $M$ point \cite{Neupert2022,Jiang_kagome}.  Our previous study suggests that, the multiband nature plays an important role in the transport properties of CsV$_3$Sb$_5$ \cite{Gan2021}. It is of great interest to investigate the generic effects  of multiband on the transport properties of all members in the  $A$V$_3$Sb$_5$ family.

In this article, we  investigate the impacts of multiband on transport properties, including Seebeck, Nernst, resistivity and Hall effect, for all members in the kagome family $A$V$_3$Sb$_5$. In the three systems, the temperature dependence of Seebeck signal and Hall resistivity switch sign at different temperatures well below the CDW transition, pointing to the important role played by multiband fermiology.  In addition, for all three members, sizable Nernst coefficient is observed,  suggesting a ubiquitous ambipolar Nernst effect in $A$V$_3$Sb$_5$.

\begin{figure}[t]
\centering
\includegraphics[width=0.4\textwidth]{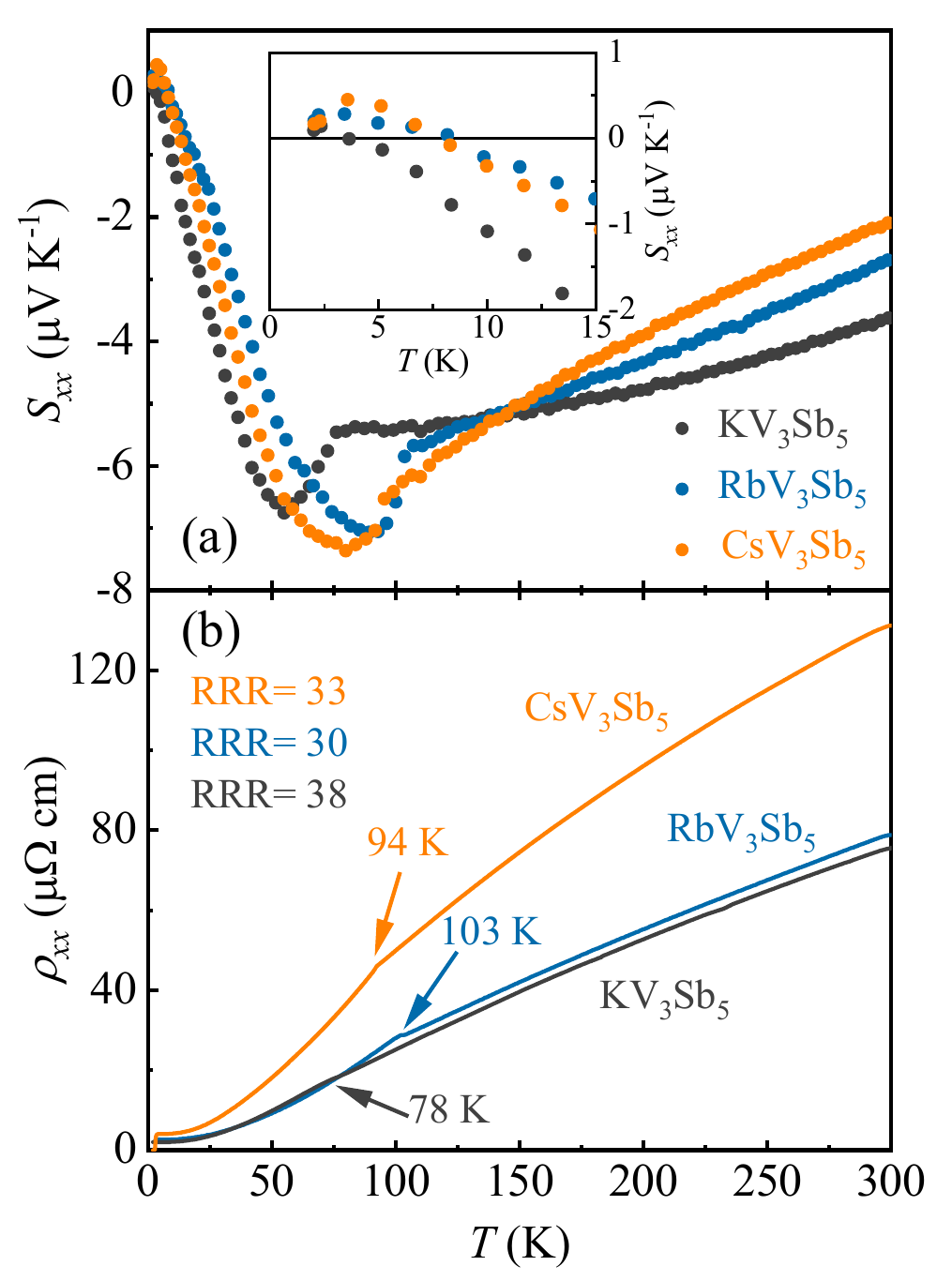}
\caption{ Temperature dependence of the in-plane Seebeck coefficient $S_{xx}$  (a), and resistivity $\rho_{xx}$ (b) measured in zero magnetic field. All three samples share similar residual resistivity ratios [RRR $=\rho_{xx}$(300 K)$/\rho_{xx}$(5 K)]. The inset in (a) is an enlarged view near the sign switching temperature of $S_{xx}$. Arrows in (b) mark out the CDW transition temperatures in $A$V$_3$Sb$_5$.}
\label{fig1}
\end{figure}

\begin{figure*}[t]
\centering
\includegraphics[scale=0.65]{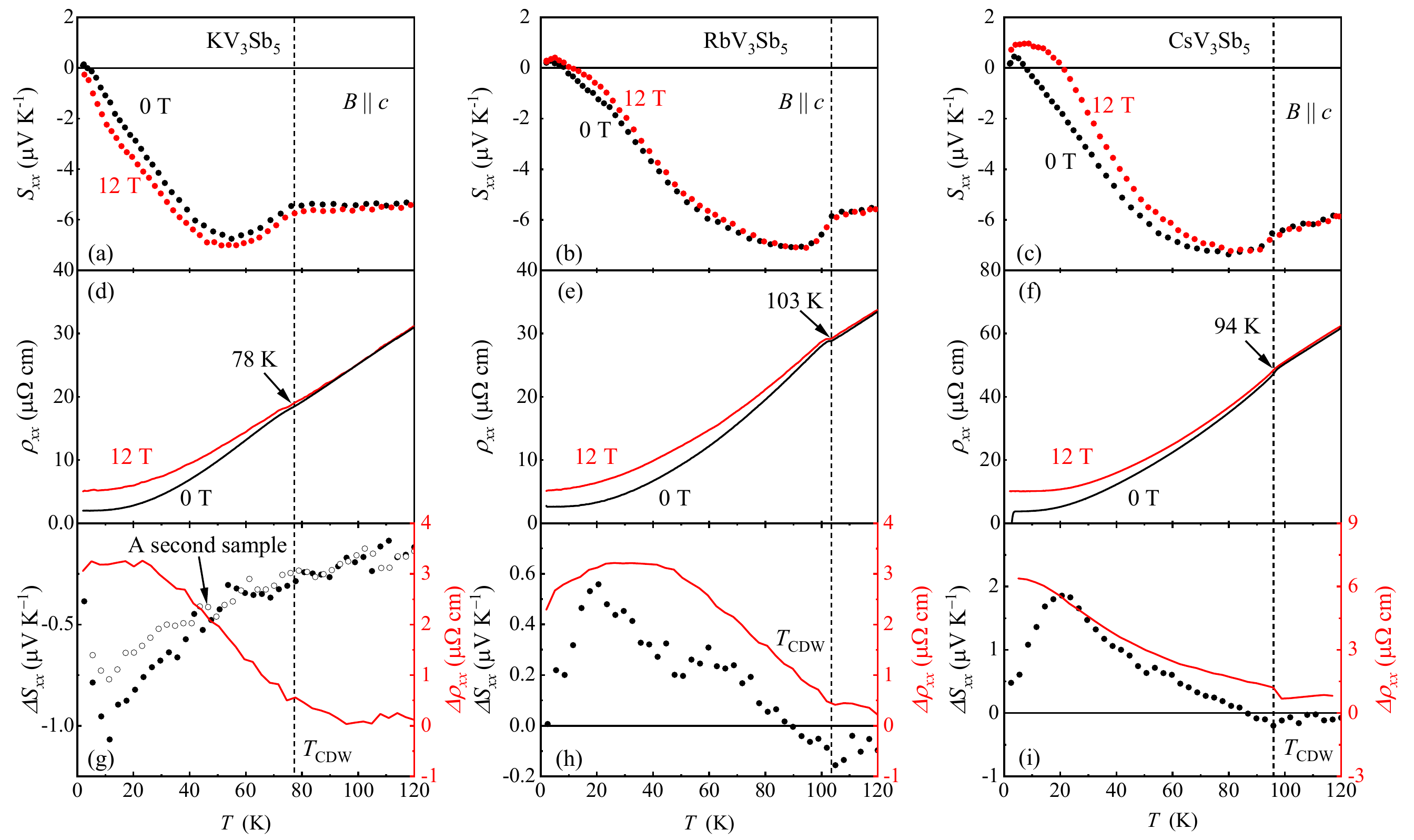}
\caption{Comparison of the in-plane Seebeck effect (a)(b)(c), and resistivity (d)(e)(f) measured in zero magnetic field and $B_z\parallel c=$ 12 T. (g)(h)(i) The difference between the data recorded in zero field and 12 T, i.e., $\Delta S_{xx}=S_{xx}$(12 T)$-S_{xx}$(0 T), and  $\Delta \rho_{xx}=\rho_{xx}$(12 T)$-\rho_{xx}$(0 T). The CDW transitions in the three systems are labeled by vertical dash lines. }
\label{fig2}
\end{figure*}

\section{Experimental Method}
Single crystals of  KV$_3$Sb$_5$, RbV$_3$Sb$_5$ and CsV$_3$Sb$_5$ were prepared by a self-flux method \cite{Li2021NodalSA}. Samples with typical dimensions of 3 $\times$ 1.5 $\times$ 0.5 mm$^3$ were used to study the in-plane transport properties. All transport measurements were carried out from 300 to 2 K in a cryostat (14 T, Oxford Instruments). A standard Hall bar geometry was employed to investigate the longitudinal and transverse resistivity simultaneously. The thermoelectric measurements were performed using one heater, two-thermometer geometry in a home-built probe. Thermal gradients and electric currents were directed within the $ab$ plane. Magnetic fields were applied along the $c$-axis.

\section{Results and Discussion}

Figure \ref{fig1} shows the results of Seebeck ($S_{xx}$), and longitudinal resistivity ($\rho_{xx}$), for  KV$_3$Sb$_5$, RbV$_3$Sb$_5$ and CsV$_3$Sb$_5$ single crystals. For better comparison, samples with similar residual resistivity ratios [RRR $=\rho_{xx}$(300 K)$/\rho_{xx}$(5 K)], were chosen for this study [see Fig. \ref{fig1}(b)]. Compared with our previous study \cite{Gan2021}, similar results are found here for CsV$_3$Sb$_5$ although a different sample with lower RRR was used.   Fermi surface reconstruction and gap opening caused by charge ordering have profound impacts on transport properties, giving rise to a sudden jump in Seebeck signal,  and a kink in resistivity, as shown in Figs. \ref{fig1}(a)(b). From these features, the charge ordering temperatures can be identified as $T_\textrm{CDW}=$ 78, 103 and 94 K for   KV$_3$Sb$_5$, RbV$_3$Sb$_5$ and CsV$_3$Sb$_5$, respectively.  The Seebeck signal keeps negative values from room temperature all the way down to about 10 K, suggesting dominating electronlike transport. Upon further cooling, a sign change in $S_{xx}$ occurs at $\sim$ 4, 8 and 7 K for KV$_3$Sb$_5$, RbV$_3$Sb$_5$ and CsV$_3$Sb$_5$, respectively [see inset in Fig. \ref{fig1}(a)].  

Such a sign switching in the temperature dependence of $S_{xx}(T)$ is a typical signature of competing electronlike and holelike excitations, i.e., multiband transport. In a multiband metal, the total Seebeck signal is considered as a weighted sum \cite{Barnard1972}, $S=\frac{\sum\sigma_i S_i}{\sum\sigma_i}$, with the Seebeck $S_i$ and conductivity $\sigma_i$ contributions from the $i$th band. A two-band approximation gives a total Seebeck coefficient as follows:
\begin{equation}
\begin{aligned}
S=\frac{\sigma_e S_e+\sigma_h S_h}{\sigma_e+\sigma_h}=\frac{\pi^{2}k_{B}^{2}T}{3e}\frac{e\mu_{h}N_{h}-e\mu_{e}N_{e}}{n_{h}e\mu_{h}+n_{e}e\mu_{e}},
\label{eq:1}
\end{aligned}
\end{equation} where $\sigma_{e(h)}=n_{e(h)}e\mu_{e(h)}$ is the electrical conductivity of electronlike (holelike) carriers with  a density $n_{e(h)}$ and a mobility $\mu_{e(h)}$. And, $S_{e(h)}=\pm\frac{\pi^{2}k_{B}^{2}T}{3e}\frac{N_{e(h)}(E_{F})}{n_{e(h)}}$, is the Seebeck contribution from electronlike (holelike) band that carries a negative (positive) sign following the free electron gas and energy-independent relaxation time approximations \cite{Mott1936,Behnia2004}. Here, $k_B$ is the Boltzman constant, $N_{e(h)}(E_F)$ is the density of states (DOS) at the Fermi level $E_F$. In this picture, the sign change of $S_{xx}(T)$ is resulted from the role switching of dominant carrier type.  The change of dominant carrier type in $A$V$_3$Sb$_5$ will be discussed in more detail in Fig. \ref{fig3}. We also note that the sudden enhancement of $S_{xx}$ by cooling across $T_\textrm{CDW}$ is likely a consequence of significantly reduced DOS ($N_h$) in the holelike band. It is known that the total DOS is suppressed at the CDW transition \cite{Tan2021}, while this effect seems to be more apparent in the holelike bands as suggested by the Seebeck results.   

In Fig. \ref{fig2}, the response of in-plane Seebeck coefficient and resistivity in the presence of an out-of-plane magnetic field is presented. At first glance, three compounds share similar response, in which sizable magneto-Seebeck and magnetoresistance mainly develop inside the CDW phase. The appearance of substantial magneto-response in the charge ordered state may be correlated with the emergence of time reversal symmetry breaking suggested by other studies \cite{Jiang2021a,Mielke2021,Yu2021b}. Upon closer inspection, dissimilarities can be easily identified, for the Seebeck effect in particular. In RbV$_3$Sb$_5$ and CsV$_3$Sb$_5$, the magneto-Seebeck effect is practically negligible above $T_\textrm{CDW}$, and the dominant role played by electronlike excitations is suppressed in a magnetic field of 12 T. For KV$_3$Sb$_5$, on the other hand, the magneto-Seebeck signal develops well above $T_\textrm{CDW}$, and the contribution of the electronlike carriers is enhanced by external magnetic fields. As shown in Figs. \ref{fig2}(g-i), the different effects can be seen more clearly by taking the difference between the Seebeck signal captured in zero magnetic field and 12 T. As displayed in Fig. \ref{fig2}(g), two samples of KV$_3$Sb$_5$ show the same behavior, implying an intrinsic behavior. As depicted in Eq. \ref{eq:1}, the total Seebeck signal is sensitive to the balance between electronlike and holelike excitations in a multiband metal. The observed different magneto-Seebeck effects are likely originated from different variations of electronic structure in the presence of magnetic fields, such as Fermi energy $E_F$, DOS at $E_F$, etc., among the three members. Although the differences might be  small, contrasting magneto-response is not unexpected since the transport properties are mainly governed by tiny pockets near the zone boundary. 

\begin{figure*}
\centering
\includegraphics[scale=0.6]{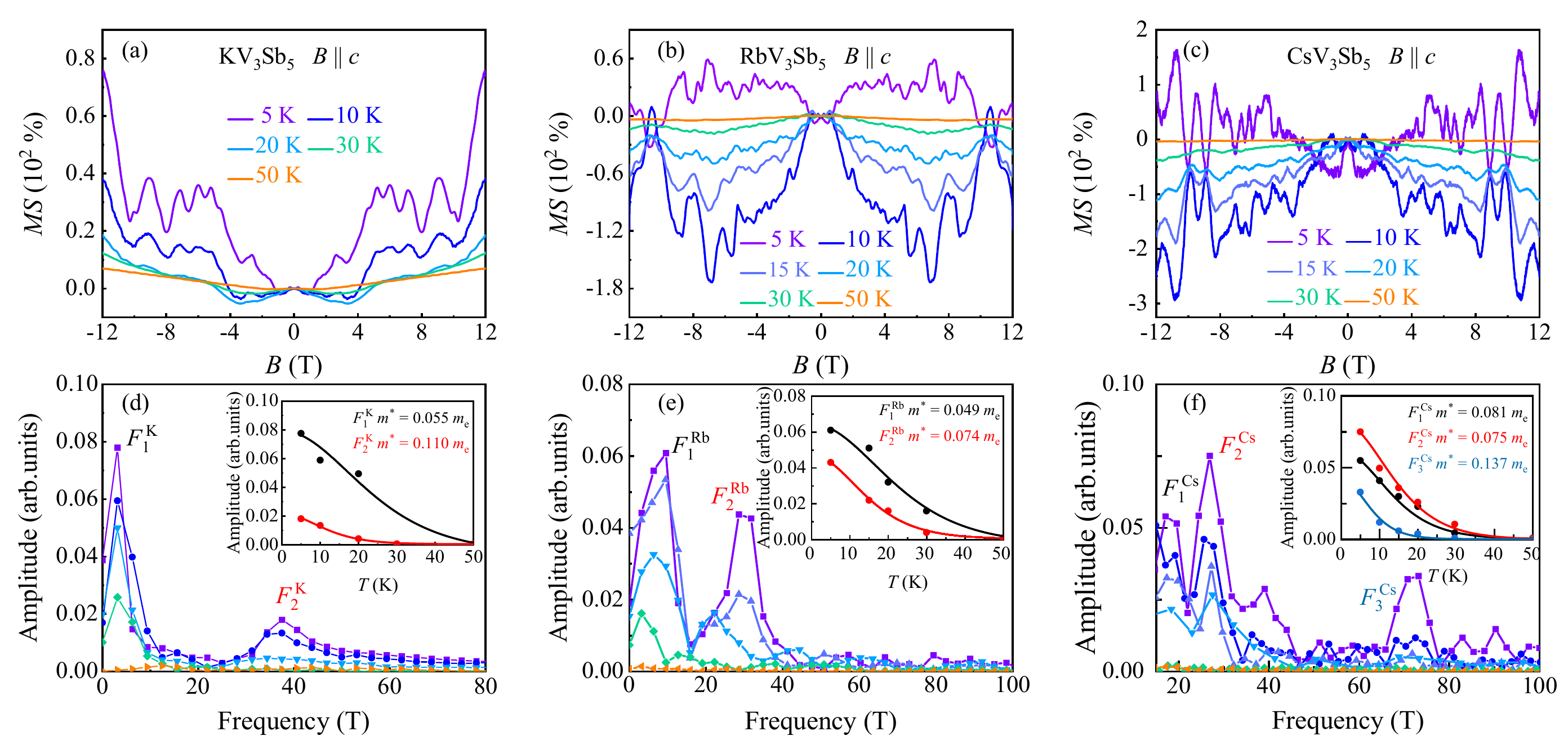}
\caption{(a-c) The magneto-Seebeck effect, $MS=[S_{xx}(B)-S_{xx}(B=0)]/S_{xx}(B=0)$, measured at fixed temperatures.  (d-f) The oscillation periods estimated from the data in (a-c) using the FFT analysis. Insets in (d-f): estimation of the effective mass ($m^*$) according to the Lifshitz-Kosevich fitting (solid lines) of the temperature dependence of the FFT peak amplitude (solid dots).  }
\label{fig:3}
\end{figure*}

The dissimilar magneto-Seebeck behaviours of the $A$V$_3$Sb$_5$ members are further seen in Figs. \ref{fig:3}(a-c). Here, the magneto-Seebeck signal was measured at selective temperatures as $MS=[S_{xx}(B)-S_{xx}(B=0)]/S_{xx}(B=0)$. Positive $MS$ values are found in KV$_3$Sb$_5$ at all temperatures studied here in magnetic fields above $\sim$5 T.   Below 30 K, negative $MS$ response appears at small magnetic fields, suggesting intricate competition between different carrier types. For RbV$_3$Sb$_5$ and CsV$_3$Sb$_5$, $MS$ shows negative values above 10 K, below which $MS$ becomes positive, in accordance with the sign change of $S_{xx}(T)$ below 10 K (see Fig. \ref{fig1} ). Besides these dissimilarities, clear quantum oscillations (QOs) are found in all three members. As shown in Figs. \ref{fig:3}(d-f), the oscillating periods were obtained using the fast Fourier transform (FFT) after subtracting a polynomial background fitted from 5 to 12 T for each compound. In all three systems, the frequencies of QOs are rather small ($<$ 100 T), agreeing qualitatively with other studies \cite{Yang2020,Yin2021,Yu2021,Gan2021}. This implies dominant roles played by the tiny Fermi pockets in the vicinity of zone boundary, since the extremal area ($A_F$) of the Fermi surface is proportional to the oscillation frequency ($F$) according to the Onsager relation $F=(\hbar/2\pi e)A_F$. The effective mass ($m^*$) of the corresponding orbits obtained using the Lifshitz-Kosevich approach is presented in the insets of Figs. \ref{fig:3}(d-f). Very small values are found in all compounds with $m^*\ll 1$, likely originated from the Dirac bands near the Fermi level \cite{Yang2020,Ortiz2021Cs}. Note that for CsV$_3$Sb$_5$, only three frequencies can be clearly identified, which are less than those found in earlier reports \cite{Ortiz2021Cs,Yu2021,Gan2021}. This discrepancy may lie in the fact that the CsV$_3$Sb$_5$ crystal measured here has a relatively low RRR value compared with samples presented in other studies.

\begin{figure*}[t]
\centering
\includegraphics[scale=0.65]{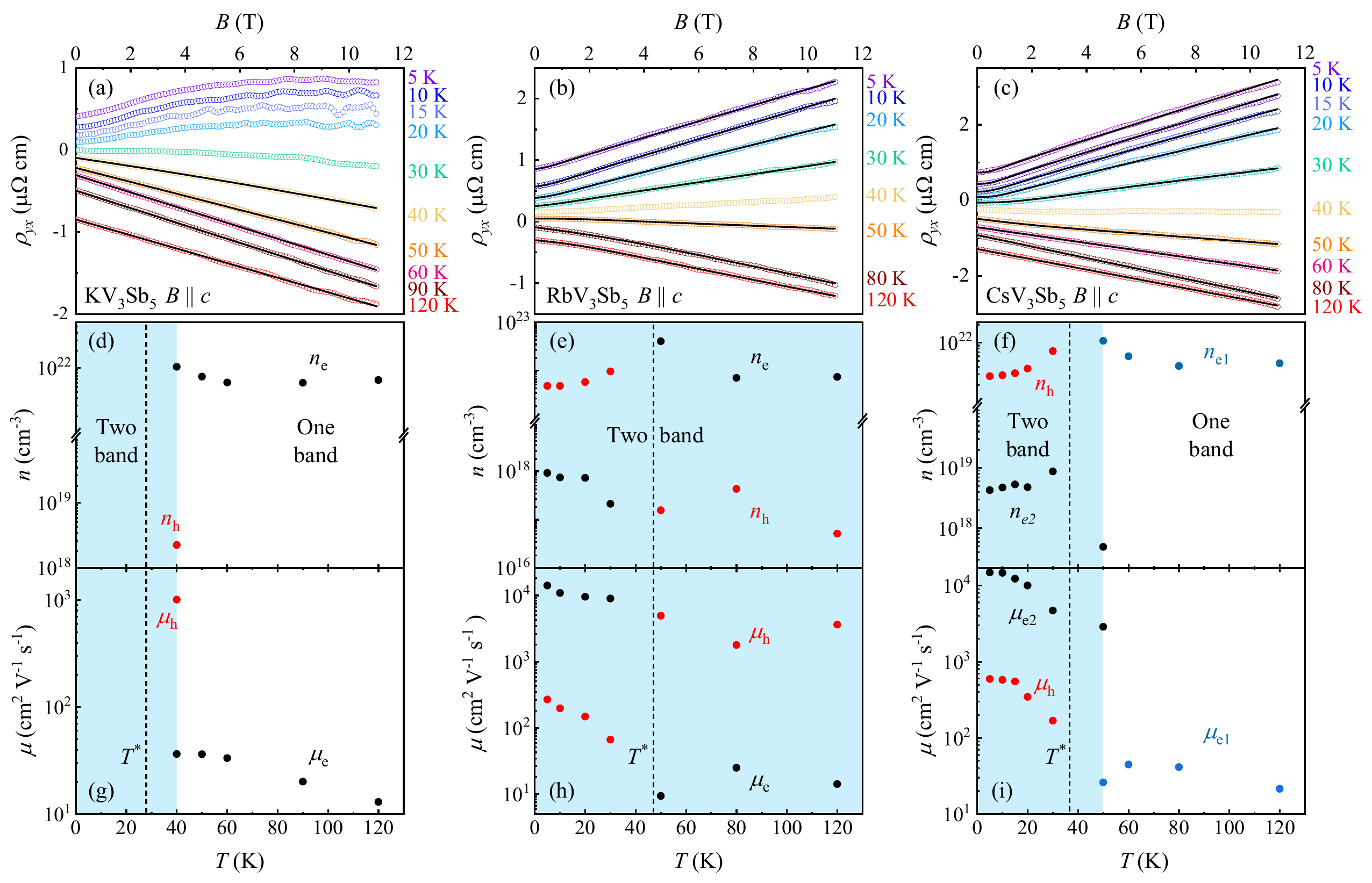}
\caption{(a-c) The Hall resistivity of $A$V$_3$Sb$_5$ captured  at selective temperatures with $B_z\parallel c$. Open circles are experimental data and solid lines are two-band modeling (see Eq. \ref{eq:2}). Vertical offsets have been applied for clarity. Temperature evolution of carrier densities (d-f)  and mobilities (g-i) obtained from the two-band analysis. Vertical dash lines tag the sign reversal temperature, $T^*$, in Hall resistivity.}
\label{fig3}
\end{figure*}

To explore the multiband transport further, we present the Hall resistivity of $A$V$_3$Sb$_5$ in Fig. \ref{fig3}. For KV$_3$Sb$_5$ and CsV$_3$Sb$_5$, the transport is dominated by one electronlike band, and the Hall resistivity, $\rho_{yx}(B)$,  scales linearly with magnetic field $B$ above 40 and 50 K, respectively. By cooling below these temperatures, nonlinear $\rho_{yx}(B)$ curves are observed, suggesting the involvement of multiband transport. The multiband effects survive up to much higher temperatures with sublinear $\rho_{yx}(B)$ appearing below 120 K in RbV$_3$Sb$_5$. The competition between electronlike and holelike carriers causes a sign switching in $\rho_{yx}$ for KV$_3$Sb$_5$, RbV$_3$Sb$_5$ and CsV$_3$Sb$_5$ at $T^*\sim$ 25, 45 and 35 K, respectively, agreeing well with earlier reports \cite{Yang2020,Yin2021,Yu2021,Gan2021}. To track down the temperature evolution of different carrier types, a two-band model is used to describe the nonlinear Hall resistivity \cite{Chambers1952}:

\begin{equation}
\begin{aligned}
\rho_{yx}(B)=\frac{B}{e}\frac{(n_{h}\mu_{h}^{2}-n_{e}\mu_{e}^{2})+\mu_{h}^{2}\mu_{e}^{2}B^{2}(n_{h}-n_{e})}{(n_{h}\mu_{h}+n_{e}\mu_{e})^{2}+\mu_{h}^{2}\mu_{e}^{2}B^{2}(n_{h}-n_{e})^{2}},
\label{eq:2}
\end{aligned}
\end{equation}
where $n_{e(h)}$ and $\mu_{e(h)}$ represent the carrier density and mobility of electronlike (holelike) excitations. Using the constrain of zero field conductivity $\sigma_{xx}=n_h e\mu_h+n_e e\mu_e$, the nonlinear Hall resistivity of  RbV$_3$Sb$_5$ and  CsV$_3$Sb$_5$ can be fitted well with the two-band model [solid lines in Figs. \ref{fig3}(b)(c)], except at 40 K where the electronlike and holelike bands almost compensate with each other. For KV$_3$Sb$_5$, however, the two-band model only works  above 30 K. At lower temperatures, involvement of AHE or more bands complicates the Hall resistivity, leading to intricate curvatures that are beyond the scope of a simple two-band analysis. Compared with earlier reports \cite{Yang2020,Yu2021,Zheng2021gate}, the AHE is less visible, possibly due to lower RRR values in the crystals studied here, since the AHE is highly sensitive to the position of Fermi level \cite{Zheng2021gate}. Evident signatures of AHE have not been reported in RbV$_3$Sb$_5$ to the best of our knowledge. Thus, the AHE varies among the $A$V$_3$Sb$_5$ series, and depends on sample quality even for the same compound. The multiband transport, on the other hand, appears to be generic in $A$V$_3$Sb$_5$ and for samples with different RRR values.   

\begin{figure*}[t]
\centering
\includegraphics[scale=0.6]{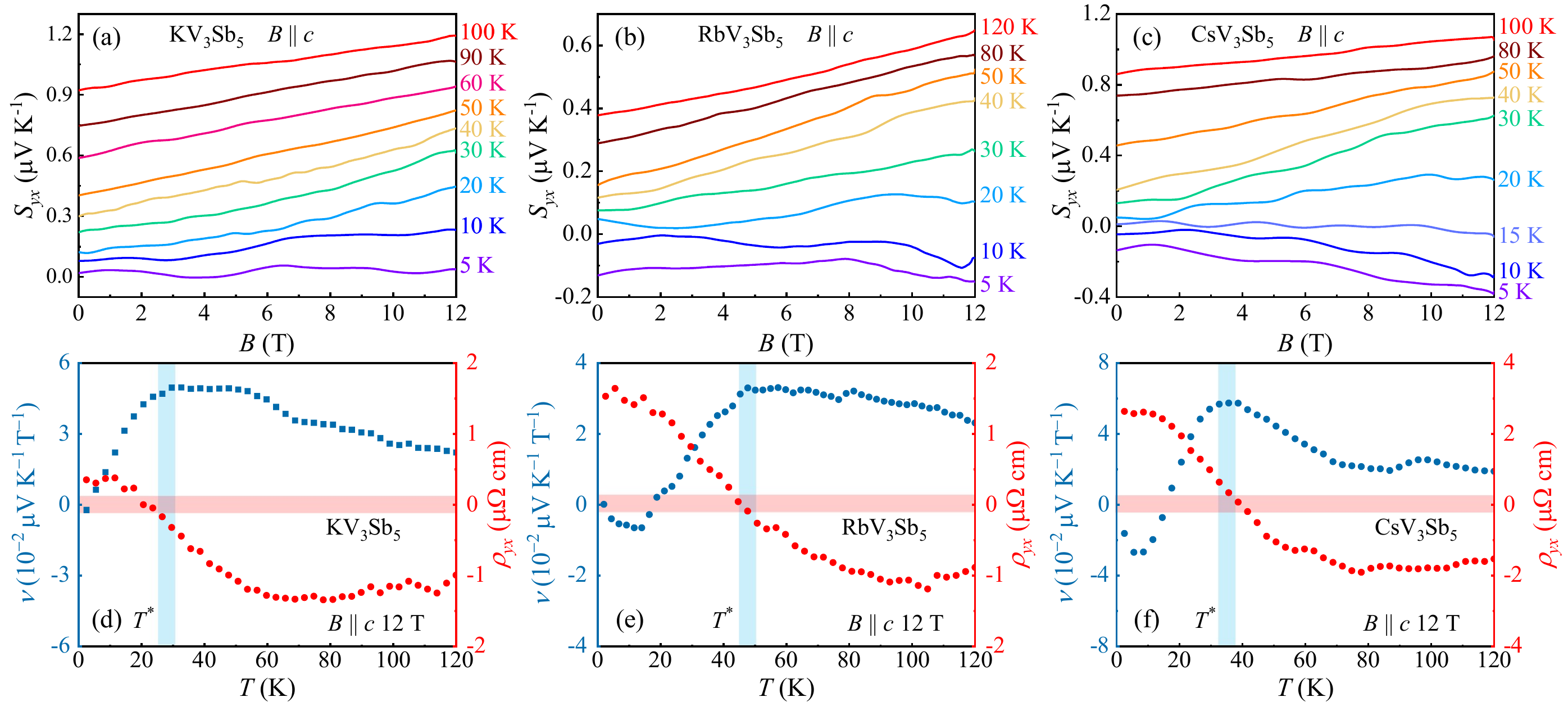}
\caption{(a-c) The Nernst signal of $A$V$_3$Sb$_5$ measured at various temperatures using $B_z \parallel c$. Curves have been shifted vertically for clarity.   (d-f) The temperature dependence of the Nernst coefficient $\nu=S_{yx}/B_z$ and Hall resistivity measured in a magnetic field $B_z=$ 12 T. The sign change temperatures of Hall resistivity, $T^*$,  at which $\nu$ reaches maximum, are highlighted by blue and red shaded stripes. }
\label{fig4}
\end{figure*}

The temperature dependence of carrier densities and mobilities extracted from the two-band analysis are displayed in Figs. \ref{fig3}(d-i). One can see that, one electron band with a density $\sim 10^{22}$ cm$^{-3}$, and a mobility $\sim 10$ cm$^2$ V$^{-1}$ s $^{-1}$, governs the Hall resistivity of KV$_3$Sb$_5$ and CsV$_3$Sb$_5$ above 40 and 50 K, respectively. For KV$_3$Sb$_5$, a hole band shows up below 40 K, which competes with the electronlike carriers and leads to a sign change in $\rho_{yx}$ below $T^*\sim$ 25 K. For CsV$_3$Sb$_5$, a second electron band emerges below 50 K before the appearance of a hole band below $T^*\sim$ 35 K, agreeing well with our previous study \cite{Gan2021}. Electronlike and holelike carriers coexist up to 120 K in RbV$_3$Sb$_5$, and a role switching between these two types of excitations occurs at $T^*\sim$ 45 K. Despite various details, the switching of dominant carrier types is seen clearly in all three systems. Note that in all three members, the sign change temperatures for Hall resistivity, $T^*$, are substantially higher than those of $S_{xx}(T)$ (see Fig. \ref{fig1}). The dominant thermoelectric contributions from electronlike carriers, $S_e$, survive down to $\sim$ 10 K. This is not unexpected from Eq. \ref{eq:1} by noticing that the mobilities of electronlike excitations are orders of magnitude higher than those of holelike counterparts below $\sim T^*$ [see Figs. \ref{fig4}(d-f)].

The transverse thermoelectric Nernst effect, $S_{yx}$, is another sensitive probe to study the balance between multiple types of carrier excitations.  In Fig. \ref{fig4}, we explore the Nernst effect in $A$V$_3$Sb$_5$.  The effects of multiband transport are again supported by the sublinear magnetic field dependence of Nernst signal, $S_{yx}(B)$, which develops below 50 K in all three compounds, as presented in Figs. \ref{fig4}(a-c). Below 20 K, combined effects of vanishing Nernst effect and QOs (see Fig. \ref{fig:3}) lead to rather fluctuating Nernst signal.   A sign change also occurs in $S_{yx}$ below 20 K for RbV$_3$Sb$_5$ and CsV$_3$Sb$_5$, which is absent for KV$_3$Sb$_5$. These different behaviors are consistent with the contrasting magneto-Seebeck effect shown in Fig. \ref{fig2}, since $S_{yx}$ has 
intimate connection with $S_{xx}$ (see Eq. \ref{eq:6}).  Figs. \ref{fig4}(d-f) compares the temperature dependence of the Hall resistivity and the Nernst signal measured in 12 T. The Hall resistivity varies continuously with temperature, and switches sign at $T^*$. Notably, the Nernst coefficient, $\nu=S_{yx}/B_z$, increases slowly upon cooling and reaches a maximum near $T^*$, below which $\nu$ decreases rapidly.   

 Generally, the Nernst signal can be described as \cite{Wangyayu2001}:

\begin{equation}
 S_{yx} =\frac{E_y}{\partial_x T}=S_{xx}\left(\frac{\alpha_{yx}}{\alpha_{xx}}-\frac{\sigma_{yx}}{\sigma_{xx}}\right),
\label{eq:6}   
\end{equation}
where $\sigma_{xx}(\sigma_{yx})$ and  $\alpha_{xx}(\alpha_{yx})$ are diagonal (off-diagonal) components of the electrical conductivity $\overline{\sigma}$ and the Peltier conductivity $\overline{\alpha}$ tensors. Here, negligible transverse thermal gradient ($\partial_{y}T\sim 0$), and isotropic diagonal response ($\sigma_{xx}=\sigma_{yy}$, $S_{xx}=S_{yy}$) have been assumed. In a one band system, vanishing Nernst signal is typically found due to Sondheimer cancellation, i.e. $\sigma_{yx}/\sigma_{xx}=\alpha_{yx}/\alpha_{xx}$, if the electrical conductivity is energy-independent  \cite{Sondheimer1948,Wangyayu2001}.  The Sondheimer cancellation does not necessarily occur in a multiband system, and finite Nernst effect can be expected \cite{Bel2003,Behnia_2009}. In the case of a two-band system, Eq. \ref{eq:6} changes to \cite{Bel2003,Behnia_2009}:
\begin{equation}
\begin{aligned}
S_{yx}=S_{xx}\left(\frac{\alpha_{yx}^{h}+\alpha_{yx}^{e}}{\alpha_{xx}^{h}+\alpha_{xx}^{e}}-\frac{\sigma_{yx}^{h}+\sigma_{yx}^{e}}{\sigma_{xx}^{h}+\sigma_{xx}^{e}}\right),
\label{eq:7}
\end{aligned}
\end{equation}
where terms containing superscripts of $h$ and $e$ correspond to the contributions from holelike and electronlike bands. Now, since the signs of $\alpha_{xx}$ and $\sigma_{yx}$ depend on the type of bands, finite Nernst response is possible. Moreover, in an ideal compensated system, one expects $\sigma_{yx}^h=-\sigma_{yx}^e$, which leads to a vanishing contribution to the Hall effect. On the other hand, $\alpha_{yx}^h$ and $\alpha_{yx}^e$ share the same sign, giving rise to an enhanced Nernst signal compared with that of a single band system.  

As discussed above, the $A$V$_3$Sb$_5$ compounds share similar multiband characteristics, which show compensated transport near $T^*$. The Sondheimer cancellation is prevented by the ambipolar flow of carriers with opposite signs, leading to  sizable Nernst signal that reaches a maximum at $T^*$. Although rarely found in metals, the ambipolar Nernst effect has been reported in another well-known CDW superconductor 2$H$-NbSe$_2$ \cite{Bel2003}. The appearance of ambipolar Nernst effect in $A$V$_3$Sb$_5$, thus, make this kagome system another prominent platform to study the ambipolar transport effects. Anomalous Nernst effect (ANE), as reported in high purity CsV$_3$Sb$_5$ crystals \cite{Chendong2021,Zhou2021}, also contributes to finite Nernst signal. However, the ANE weakens gradually upon warming, and disappears above $T^*$. Thus, the enhanced Nernst amplitude near $T^*$ observed here is likely dominated by ambipolar transport, while ANE only comes into play at lower temperatures \cite{Zhou2021}.

The nature of the observed peculiar sign reversal in the temperature dependence of the Hall coefficient well below $T_\textrm{CDW}$ is still elusive.  As suggested by R. Bel \textit{et al}., radical changes in the mean-free path of charge carriers could be responsible for the sign switching of the Hall coefficient  below the CDW transition in 2$H$-NbSe$_2$ \cite{Bel2003}. Similar physics may happen in $A$V$_3$Sb$_5$, where the mean-free path of holelike excitations is enhanced substantially below $T^*$, compared with that of the electronlike counterpart. On the other hand, an electronic nematic transition is observed around $T^*\sim$ 35 K in CsV$_3$Sb$_5$ by means of elastoresistivity, nuclear magnetic resonance and scanning tunneling microscopy experiments \cite{Nie2022}. Two additional Raman modes also appear below $\sim$ 30 K \cite{Li2021}, and the muon relaxation rates show a deep increase below $\sim$ 30 K \cite{Yu2021b},  for CsV$_3$Sb$_5$. In KV$_3$Sb$_5$, a minimum is observed around 30 K in the Knight shift experiments \cite{Mielke2021}. These results suggest the existence of possible symmetry breaking around $T^*$ . It is also of great interest to study the connection between the sign reversal in the temperature dependence of the Hall effect and possible additional symmetry breaking effects deep inside the CDW phase.       

\section{Conclusions}

In summary, we have investigated multiband transport properties in $A$V$_3$Sb$_5$. In spite of minor differences, few generic features are found in the CDW phase for all three members, including sign changes in the temperature dependence of the Seebeck and Hall effects, together with sublinear magnetic field dependence of the Hall and Nernst effects. Moreover, the ambipolar flow of compensated charge carriers with opposite signs gives rise to sizable Nernst signal. These results suggest that the multiband transport effects are generic features in $A$V$_3$Sb$_5$, and that this kagome family is another novel platform to study the interplay between CDW and ambipolar transport effects in metallic systems.

\section{Acknowledgments}
We thank Hengxin Tan, Zhiwei Wang for stimulating discussions. We thank Guiwen Wang and Yan Liu at the Analytical and Testing Center of
Chongqing University for technical support. This work has been supported
by National Natural Science Foundation of China (Grant No. 11904040), Chongqing Research Program of Basic Research and Frontier
Technology, China (Grant No. cstc2020jcyj-msxmX0263). Y. Chai acknowledges
the support by National Natural Science Foundation of China (Grant
Nos. 11674384, 11974065). Y. Guo acknowledges the support by the Major Research Plan of the National Natural Science Foundation
of China (No. 92065201). A. Wang acknowledges the support by National
Natural Science Foundation of China (Grant No. 12004056).

\bibliographystyle{apsrev4-1}

\end{document}